\documentclass[]{spie}  
\usepackage[]{graphicx}
\usepackage[]{multicol}
\usepackage[latin1]{inputenc}

\title{VSI: a milli-arcsec spectro-imager for the VLTI} 

\author{F. Malbet\supit{a}, P.Y. Kern\supit{a}, J.-P. Berger\supit{a},
  L. Jocou\supit{a}, P. Garcia\supit{b}, D. Buscher\supit{c},
  K. Rousselet-Perraut\supit{a}, G. Weigelt\supit{d}, M. Gai\supit{e},
  J. Surdej\supit{f}, J. Hron\supit{g}, R. Neuhäuser\supit{h},   
  E. Le Coarer\supit{a}, P.R. Labeye\supit{i},
  J. Le Bouquin\supit{a}, M. Benisty\supit{a}, E. Herwats\supit{a} 
\skiplinehalf
\supit{a}Laboratoire d'Astrophysique de Grenoble (LAOG, France); \\
\supit{b} Centro de Astrofísica da Universidade do Porto (CAUP,
Portugal);\\
\supit{c} Cavendish Laboratory, University of Cambridge (United Kingdom);\\
\supit{d} Max-Planck Institut für Radioastronomie in Bonn (MPIfR, Germany);\\
\supit{e} Istituto Nazionale di Astrofisica - Osservatorio Astronomico
di Torino (INAF-OATo, Italy);\\ 
\supit{f} Institut d'Astrophysique et de Géophysique, Université de Liège (IAGL, Belgium);\\
\supit{g} Institut für Astronomie, Universität Wien (IfA, Austria);\\
\supit{h} Astrophysikalisches Institut und Universitäts-Sternwarte (AIU Jena,
    Germany);\\
\supit{i} CEA-LETI (France).
}

\authorinfo{Further author information: (Send correspondence to
  F.~Malbet)\\
  E-mail: fabien.malbet@obs.ujf-grenoble.fr}

\begin{document} 
  \maketitle 

\begin{abstract}
  VLTi Spectro-Imager (VSI) is a proposition for a second
  generation VLTI instrument which is aimed at providing the ESO
  community with the capability of performing image synthesis at
  milli-arcsecond angular resolution. VSI provides the VLTI with an
  instrument able to combine 4 telescopes in a baseline version and
  optionally up to 6 telescopes in the near-infrared spectral domain
  with moderate to high spectral resolution. The instrument contains
  its own fringe tracker in order to relax the constraints onto the
  VLTI infrastructure. VSI will do imaging at the milli-arcsecond
  scale with spectral resolution of: a) the close environments of
  young stars probing the initial conditions for planet formation; b)
  the surfaces of stars; c) the environment of evolved stars, stellar
  remnants and stellar winds, and d) the central region of active
  galactic nuclei and supermassive black holes.  The science cases
  allowed us to specify the astrophysical requirements of the
  instrument and to define the necessary studies of the science group
  for phase A.
\end{abstract}



\section{Introduction}
\label{sect:intro}  

VSI is proposed as second generation VLTI instrument providing the ESO
community with the capability of performing image synthesis at
milli-arcsecond angular resolution.  Image synthesis is the standard
operation of radio and (sub-)mm interferometers. VSI is the result of
the merging of two previous concept studies,
VITRUV\cite{2005astro.ph..7233M} and BOBCAT\cite{bobcat}.  They were
previously presented at the ESO workshop in 2005. VSI provides the
VLTI with an instrument able to combine 4 telescopes in a baseline
version and optionally up to 6 telescopes in the near-infrared
spectral domain with moderate to high spectral resolution. The
instrument contains its own fringe tracker in order to relax the
constraints onto the VLTI infrastructure.


A preliminary system analysis of VSI, allowed us to clarify the high
level specifications of the system, the external constraints and to
perform a functional analysis.  In particular, the instrument was
separated into 14 functions, the context of PRIMA was addressed and
the system tasks for the required phase-A study were defined.  Two
solutions for the science beam combiner were identified one based on
integrated optics and another on bulk optics.  These two solutions are
inherited from the two concepts merged. One of the goals of the Phase
A study is to define which of them will be used by VSI.  VSI has an
internal fringe tracker which relaxes the constraints on the VLTI
interfaces by allowing to servo optical path length differences of the
input beams to the required level.

The paper describes the science cases, the preliminary system study
and presents the possible concepts.

\section{Science cases}

VSI concept is a general purpose instrument aimed at exploiting the
full capability of the VLTI infrastructure including the faint science
space enabled by PRIMA. VSI is up to 5 times faster than current
interferometric instrumentation (AMBER) because it combines up to 6
telescopes.  The wavelength range is JHK.  Three spectral resolutions
are available $\sim$100, $\sim$1000 and $\sim$10000.  The dynamic
range of the reconstructed images is 10-100 with a goal of 100-1000.
There is a goal of retaining polarization information.  The current
science cases definition methodology was to concentrate in a few
fields where VSI can make a substantial contribution, without being
exhaustive.

\subsection{The formation of stars and planets} 

The early evolution of stars and the initial conditions for planet
formation are determined by the interplay of accretion and outflow
processes. Due to the small spatial scales where these processes engines
actuate, very little is known about the actual physical and chemical
mechanisms at work.  Interferometric imaging at 1~mas
(milli-arcsecond) will directly probe the regions responsible for the
bulk of continuum emission excess from these objects therefore
constraining the currently highly degenerate models for the spectral
energy distribution.  In the emission lines a variety of processes
will be probed, in particular outflow and accretion magnetospheres.
The inner few AUs of evolved planetary systems will also be studied,
providing additional information on their formation and evolution
processes, as well as on the physics of extrasolar planets.

\subsection{Imaging stellar surfaces} 

Optical imaging instruments are a powerful means to resolve stellar
features at the generally patchy surfaces of stars throughout the HR
diagram.  Optical interferometry has already proved its ability to
derive surface structure parameters such as limb darkening or other
atmosphere parameters. VSI, as an imaging device, is of strong
interest to study various specific features as vertical and horizontal
temperature profiles, abundance inhomogeneities and detect their
variability as the star rotates and pulsates. This will provide
important keys to address stellar activity processes, mass-loss
events, magneto-hydrodynamic mechanisms, pulsation and stellar
evolution.

\subsection{Evolved stars, stellar remnants and stellar winds}   

HST and ground-based observations revealed that the geometry of young
and evolved PNe and related objects (e.g.\ nebulae around symbiotic
stars) show an incredible variety of elliptical, bi-polar,
multi-polar, point-symmetrical, and highly collimated (including jets)
structures. The proposed mechanisms explaining the observed geometries
(disks, MHD collimation and binarity) can only be tested by
interferometric imaging at 1~mas resolution.

Extreme cases of evolved stars are stellar black holes.  In
microquasars the stellar black-hole accretes mass from a donor. The
interest of these systems lies in the small spatial scales and high
multi-wavelength variability.  Milli-arcsecond imaging in the NIR will
allow to disentangle of dust from jet synchrotron emission, compare the
observed morphology with radio maps and correlate it with the variable
X-ray spectral states.

\subsection{Active galactic nuclei and supermassive black holes}

AGN are complex systems composed of different interacting parts
powered by accretion onto the central supermassive black hole. The
imaging capability will allow study of the geometry and dust
composition of the obscuring torus, testing radiative transfer models.
Milli-arcsecond resolution imaging will allow to probe the collimation
at the base of the jet and the energy distribution of emitting
particles.  Supermassive black holes masses in nearby (active)
galaxies can be securely measured and it will be possible to detect
general relativistic effects for the stellar orbits closer to the
galactic center black hole.  The wavelength-dependent
differential-phase variation of broad emission lines will provide
strong constraints to the size and geometry of the Broad Line Region.
It will then be possible to establish a secure size-luminosity
relation for the BLR, a fundamental ingredient to measure supermassive
black hole masses at high redshift.

\subsection{Comparison with existing instrument capabilities}

The general purpose spectral range and resolution of VSI combined with
its 2-5 times higher efficiency makes it a natural successor to AMBER.
This second generation VLTI instrument will fully exploit the faint
science parameter space opened up by PRIMA. VSI imaging complements
spectroscopy with VLT adaptive optics angular resolution (NACO, SINFONI),
as well as spectro-imaging with ALMA. It will provide a zoom-in
capability on parts of a target that remain unresolved by AO and/or
ALMA, which is often critical for the interpretation of the large
scale imaging and spectroscopy.

Once equipped with VSI, the VLTI will be more capable than
any competing near-infrared imaging array. For example, the VLTI will
be much more sensitive than NPOI, and will provide better image
fidelity than CHARA and the Keck Interferometer, thanks to its
relocatable ATs. With regard to future arrays, the augmented VLTI will
be more than competitive with the six-telescope MROI Phase I since the
inclusion of the larger diameter UTs will provide better sensitivity.

\section{System analysis}

We have carried out a system study aimed at defining a preliminary
conceptual design for a multipurpose near-infrared spectro-imager for
the VLTI. These studies, matching as much as possible the science case
requirements, have raised several mandatory questions that will have
to be addressed during the phase A study.  The simple idea behind this
work is to provide the astronomical community with an efficient
spectro imager able to fulfill a broad science program. One additional
important constraint has been and will be continuously taken into
consideration: the requirement that VSI should be an \textbf{easy to
  maintain instrument}.  This system study has taken benefit of the
extensive experience of members of the consortia past experience in
previous successful facilities and instruments (e.g. COAST,
AMBER/VLTI, IONIC/IOTA and IONIC/VLTI).

In the initial VITRUV study the fringe tracking instrument was not
included, BOBCAT study included it. It appears that the capability of
VSI to carry out its science program depends heavily on the
VLTI ability to cophase its telescopes.  We have therefore considered
that a phase-A study should include an analysis of what is expected
as far as VLTI cophasing is concerned. 

\subsection{High level specification}

Although the initial work done by the science group has allowed us to
better constrain what should be the range of performance of
VSI further work is needed, the science case phase-A study will
have to answer the following questions that will directly impact the
instrument observing modes.

\begin{multicols}{2}
\begin{enumerate}
\setlength{\itemsep}{0pt}
\item expected image complexity;
\item dynamic range;
\item spectral coverage and dispersion requirement;
\item limiting magnitude;
\item field of view;
\item time resolution (i.e duration to obtain an image);
\end{enumerate}
\end{multicols}
This in turn will allow the system study to define high-level
technical requirements. These requirements will concern mainly (i)
what is the level of (u,v) coverage expected to access to a given
complexity; (ii) what is the level of visibility and phase
(closure-phase) accuracy expected; (iii) what is the level of array
cophasing accuracy that is expected.  At the time of the study we
consider that the VSI will be able to combine four telescopes as a
basic requirement but should include a detailed description of its
ability to combine six telescopes (goal).

\paragraph{VLTI infrastructure}

VLTI can provide 4 UT telescopes and 4 AT telescopes. Six delay lines
are available. Our starting point is considering that VSI
should be able to manage the combination of four telescopes. An
additional mode where VSI can combine six telescopes
to take full benefit of the current infrastructure will be also
considered. This latter should not be taken lightly since the impact
on imaging capability of switching from 4T to 6T is
considerable. VLTI offers the possibility of phase referencing thanks to
the PRIMA mode. Two Star Separator Systems (STS) are already
available but two additional ones are foreseen.

\paragraph{The imaging paradigm}

VSI is intended to be an imager with spectral resolution. The final
astronomical product will therefore be an image at each spectral
channel.  VSI will be able to measure sufficient visibilities and
phase information to permit model-independent image reconstruction.
This puts a strong constraint on the number of $(u,v)$ points for
which one needs to obtain visibility, closure phase and
differential-phase measurements. In particular increasing the number
of telescopes has an immediate impact on the amount of phase
information that can be retrieved through the use of closure phases
quantities.  The importance of retrieving phase information is
considerable and three options arise: (i) using a second source as a
phase reference; (ii) using the closure phase technique; (iii) in
peculiar cases using spectral differential phases;

The closure-phase technique allows us to retrieve atmosphere-free
phase information.  In a standard interferometer absolute phase
information is lost due to multiple wave-front perturbations,
(optomechanical instability, atmospheric piston). Using combination of
at least three telescopes allows to extract the so-called
closure-phases by summing up each baseline interferogram phases. This
summation cancels out all the parasitic phase perturbation and
produces the closure phase. Using an increasing number of telescopes
allows to reduce the amount of phase information difference between
phase and closure phases.
\begin{figure}[t]
  \centering
  \parbox{0.4\hsize}{\includegraphics[width=\hsize]{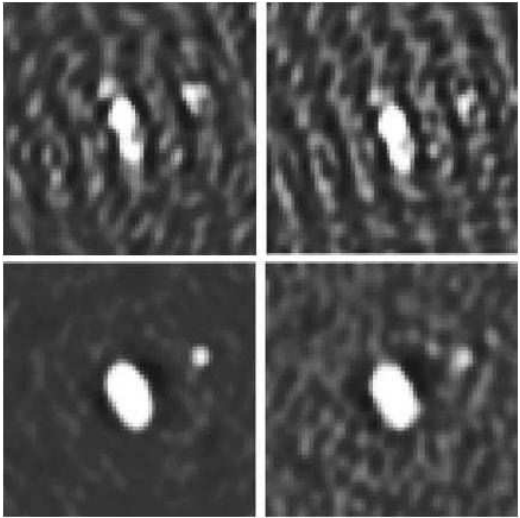}}\hfill
  \parbox{0.55\hsize}{\caption{\label{fig:4t6t}: Image reconstruction
      from simulated high-SNR data of an elliptical star with a
      companion which is approximately 3.4 magnitudes fainter than the
      primary. The upper images are reconstructed from simulated data
      using the beams from only 4 telescopes (i.e 6 instantaneous
      baselines), while the lower images are reconstructed from an
      array of 6 telescopes (i.e 15 instantaneous baselines). In each
      case an earth-rotation synthesis of 6 hours duration was
      simulated. The leftmost images are reconstructed from
      uncorrupted phase data, simulating data from a phase reference
      system, while the rightmost images are reconstructed from
      closure-phase data. All images have the same greyscale levels.
      It can be seen that the difference between images reconstructed
      from 4-telescope data and those reconstructed from 6-telescope
      data is far greater than the difference between images
      constructed from phase-referenced data and from closure phase
      data.}  }
\end{figure}
Although extracting phase information from closure phase measurements
is a tough work a considerable amount of research to find efficient
algorithms has been carried (radio) and is still underway. We believe
that the considerable relaxation of instrumental/operation constraints
introduced by the use of closure phase quantities instead of phases is
worth the already successful effort to find numerical ways to reduce
the degeneracy when phases are extracted from closure phases.

Of course the possibility of accessing directly to phase measurements
is very interesting and the phase-A study will include a detailed
study of the impact of phase vs. closure phase measurements on the
final image reconstruction capability. However we can anticipate that
the gain in terms of image quality due to the use of phase instead of
closure phase might not be as important as the gain due to the
increase in the number of telescopes. See Fig.~\ref{fig:4t6t} and
caption for an illustration. It should be remembered that imaging with
the VLTI, which requires a good level of cophasing will have an
important impact on VLTI operation.

\paragraph{Image complexity}

Image complexity depends directly on the ability to cover the (u,v)
plane with a great number of independent visibility, phase/closure
phases measurements.

\paragraph{Dynamic range}

The science case has determined that imaging with a dynamic range of
100 should allow to fulfill a significant fraction program. The
conditions at which a dynamic range of 1000 can be reached should be studied.

\paragraph{Spectral coverage and dispersion requirement}

The science case study has defined operation at J, H and K bands
with two to three spectral resolution spanning the [100,10000] domain
as the minimum configuration. Some programs in the different legacy
surveys may require R$\approx 30000$.  

\paragraph{Limiting magnitude}

Currently the faintest objects contained in the science case have
magnitudes of $H,K\approx14$. 

\paragraph{Field of view}

The science case is mainly focused towards ``compact'' sources i.e
sources that are not individually resolved by the UTs at their
diffraction limit. Currently most of the science programs require a
field of view no bigger than 0.2''. During phase A the science group
will have to define: (i) if the sources are indeed unresolved by
individual apertures; (ii) the interest in increasing the field of
VSI, so that it is able to map extended structures (in the sense of
bigger than individual diffraction fields of view).

\paragraph{Time resolution}

VSI will be able to provide an image within one night. However
the phase A science study should take into consideration the
impossibility to move the telescope configuration during the night
(especially with four telescopes) and should evaluate the impact on
science of observing the same object with different configurations
obtained at different epochs. Currently preliminary science cases have
pointed out objects with intrinsic variability ranging from  1 hour to
1 month.

\subsection{VSI external constraints}

\paragraph{Atmospheric refraction and dispersion}

Since stellar light passes through a prism of atmosphere,
the different wavelengths are refracted with different
angles that depend upon zenithal angle. These refraction
angles significantly vary from a spectral band to another,
and even through a spectral band, as detailed and shown
in the AMBER studies. So, the resulting
image spots in the J and H bands are spectrally dispersed
and thus appear elongated of an order of a few Airy disks.
In the particular case of a single-mode instrument
this leads to a coupling efficiency degradation at the extreme
wavelengths of these bands. Another consequence lies in the
fact that the angle between the beam direction provided by
the adaptive optics device to the image sensor and the
direction of the actual observing wavelength varies with
time, inducing a variation of the coupling efficiency.

\paragraph{Atmospheric dispersion}

Since stellar light does not follow the same horizontal
optical path in the air, an optical path difference (OPD)
proportional to the interferometric baseline and to the
projection of the zenithal angle on the meridian plane exists.
As computed in the AMBER studies, this
effect is rather negligible with the exception of long baselines and
at low spectral resolution. In these cases, visibility loss can reach
several percents but quickly decrease with spectral
resolution. Moreover the corresponding bias can be well modeled.

\paragraph{Field of View}

Whatever the telescopes, the unvigneted field of view (FOV)
at the instrument input has a diameter of 2". The ESO facility
PRIMA allows a dual-feed mode with two FOV of 2", separated
by an Airy disk at least and picked up anyway on the global
Coud\'e field of 2'.



\paragraph{Beam optical quality}

The typical tip-tilt error budget provides errors smaller than 21 mas
rms on the sky with the UTs, and than 30 mas rms on the sky with the
ATs, in the single feed mode.  Based on the current experience with
AMBER it seems reasonable to reassess the tip/tilt performances at
VLTI in order to determine the need for an additional module
integrated to VSI that would allow additional tip/tilt and/or higher
order modes adjustment.

\paragraph{Optical path differences}

The VLTI has been designed to be intrinsically stable and,
without fringe tracking, internal VLTI OPD fluctuations of
338 nm in K band are expected for an exposure time of 48 ms.

\paragraph{Polarization}

The VLTI has been designed to minimize the differential
polarization effects thanks to symmetric optical trains.
Nevertheless due to multiple reflections in each arm,
various optical coatings, aging of these coatings, etc.
residual polarization effects as partial polarization and
phase shifts between the two perpendicular directions of
linear polarization and/or between two different interferometric
arms remain. 

\paragraph{VLTI throughput}

The VLTI throughput with the UTs and the ATs equals from
20\% up to 35\% over the J, H and K bands, according to
the wavelength and the optical configuration.

\subsection{Functional analysis}


We have carried out a sub-system breakdown in order to define for each
of the instrument functions what were the studies that had to be
addressed during the phase A study.  At the current level of the study
we have not merged any of the two different concepts (BOBCAT and
VITRUV) but we can reasonably agree on the subsystem breakdown.  As we
will see in next section the preliminary conceptual designs arising
from that are quite different and the phase A study will have to
define the best concept. The VSI 14 subfunctions are made of:
\begin{multicols}{2}
  \begin{enumerate}
\setlength{\itemsep}{0pt}
\item atmospheric dispersion compensator;
\item spatial filtering;
\item wavefront correction;
\item fringe tracking;
\item optical path scanner;
\item beam injection;
\item beam combiner;
\item polarization control;
\item spectral dispersion;
\item detector;
\item data processing;
\item calibration and alignment tools;.
\item control module;
\item image reconstruction.
\end{enumerate}
\end{multicols}
It should be stated that some of the previous subsystems might be
irrelevant depending on the final beam combination concept
adopted. Also the fringe tracking instrument should be seen as an
instrument by itself.

\section{Conceptual designs}

This section presents the technical solutions selected for the
sub-systems of VSI. 

\subsection{Integrated optics}
\label{sect:conceptIO}

The combination proposed here is realized with Integrated Optics
technology. This one is well suitable to the combination of four
telescopes or more and the performances in terms of stability and
efficiency have been demonstrated at several occasions on
VLTI\cite{2004A&A...424..719L,2005astro.ph.12544L} and
IOTA\cite{2004ApJ...602L..57M,2005AJ....130..246K}. 
\begin{figure}[t]
\centering
\includegraphics[width=0.55\hsize]{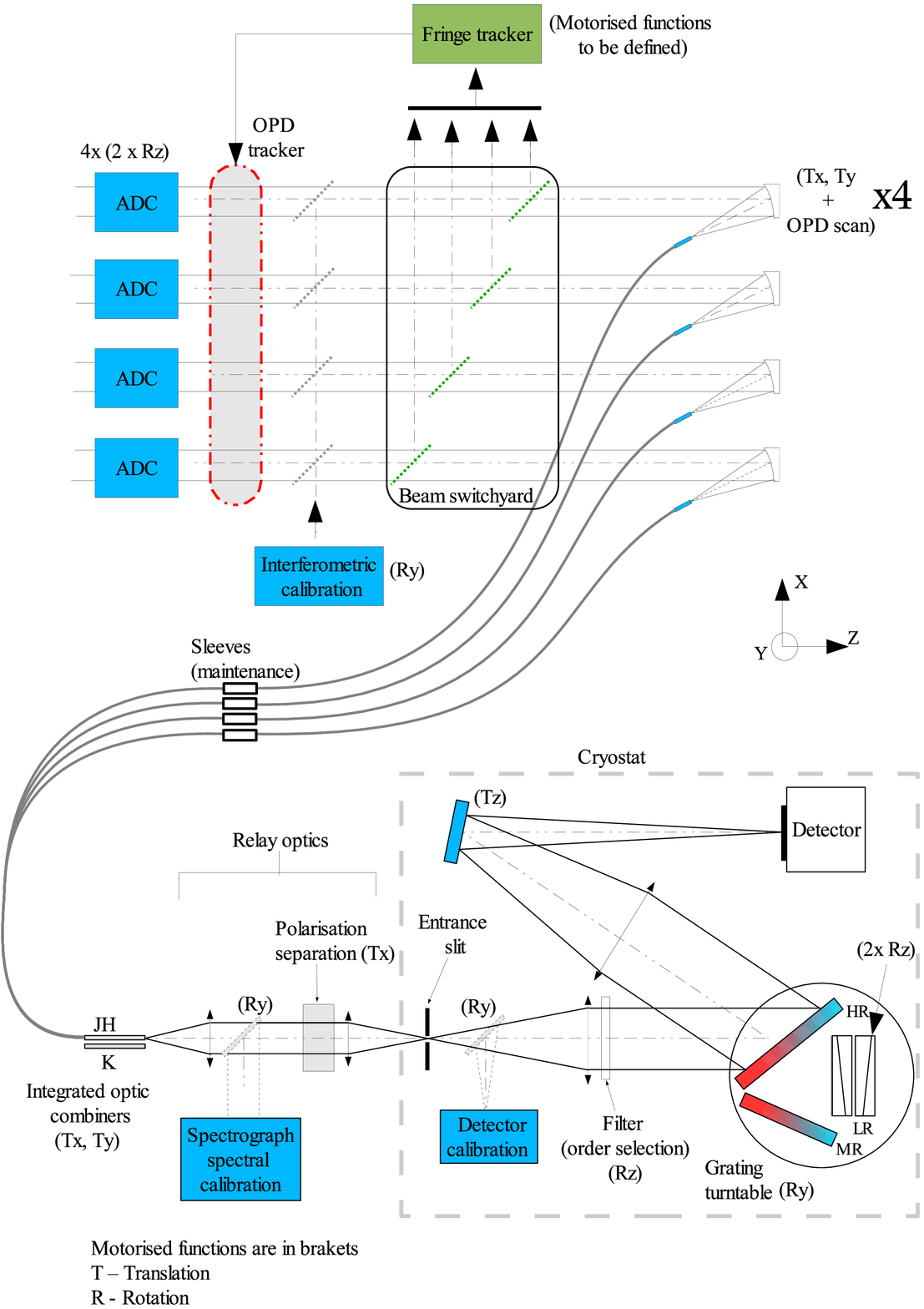}
\caption{Schematic view of the beam combiner concept}
\label{fig:conceptIO}
\end{figure}
Figure \ref{fig:conceptIO} describes a conceptual study of the
science beam combiner derived from our system analysis. This last is
composed of the following subsystems:
\begin{multicols}{2}
\begin{itemize}
\setlength{\itemsep}{0pt}
\item Atmospheric dispersion compensator modules;
\item Alignment/calibration module;
\item Beam switchyard/Fringe tracker;
\item Beam injection module;
\item Spatial filtering;
\item Beam combination module;
\item Polarization control module;
\item Spectrograph;
\item Spectral calibration module;
\item Detector;
\item Control software;
\end{itemize}
\end{multicols}
The control Software and data reduction aspects are not represented in
the figure but their functions are approached in the next sections.
As it is proposed in the system analysis, the combination of 4 VLTI
beams is our baseline while the combination of 6 or 8 beams is
proposed as an additional option. We have considered IO technology as
our solution to combine the 4 to 6/8 VLTI beams. This choice is based
on LAOG experience on this type of combination which has been
successfully exploited at VLTI (2-way beam) and IOTA (2,3-way beam
combiners). The main advantages of this technology are:
\begin{itemize}\setlength{\itemsep}{0pt}
\item possibility to integrate on single chip almost all combination schemes;
\item it is easy to implement the modal filtering function associated
  to photometric calibration which has proved to be a key element in
  the improvement of the visibility accuracy;
\item the very compact size of the combining chip allows a remarkable
  stability of phase properties and a small combining footprint;
\item easy to install since only the alignment with the spectrograph
  has to be ensured;
\end{itemize}
Each beam combiner is made of an integrated optics chip connected to a
fiber V-groove.  At the present state of the study, the capability to
use of only one chip to recombine both the J and H bands has not been
demonstrated. Our baseline is consequently to ensure the combination
of the 4 VLTI beams with three components, one component per spectral
band. In the global study of the combiner, we foresee to study however
the use only one IO component for both J and H bands.

Among all the beam combination concepts, our system study has pointed
out that a four way pairwise ABCD beam combiner was the best
compromise as far as signal to noise and biases are concerned.  Our
industrial partner LETI has designed such a combination concept. 

\subsection{Bulk optics}
\label{sect:conceptBO}

The bulk-optics (BO) option is an alternative to the integrated-optics
(IO) science combiner which offers a number of potential advantages.
Chief amongst these is the combination of high photon throughput
($>$96\%) and high fringe contrast ($>$95\%) which has been
demonstrated in working prototypes.  

The BO science combiner has the same basic functionality as the IO
combiner. The main difference in implementation is the use of
free-space optics rather than guided-wave optics for producing the
interference patterns. A number of beam combination modes are possible
within the BO concept, including image-plane and pupil-plane
combination, as are a number of spatial filtering modes such as
pinhole and fibre filtering. For simplicity we describe here mainly
the baseline option, which uses pupil-plane beam combination and fibre
spatial filtering.
\begin{figure}
  \centering
  \includegraphics[width=0.55\textwidth]{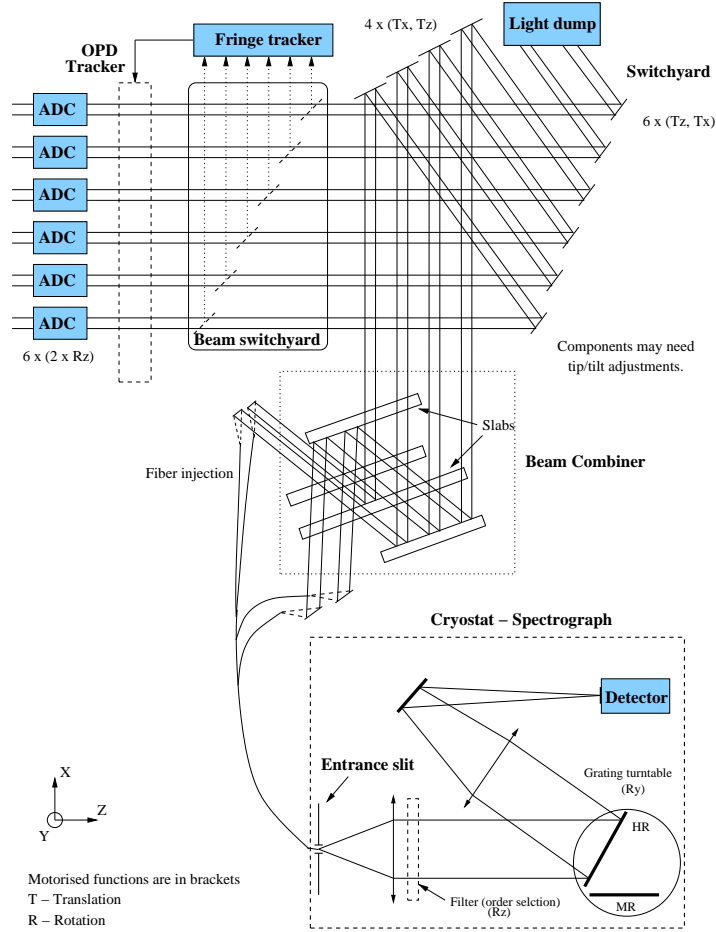}
  \caption{Schematic layout for the bulk-optics combiner. The layout
    for an instrument accepting 6 input beams is shown, including a
    switchyard for selecting a subset of the beams to feed into a
    4-way combiner.}
  \label{fig:bulkschematic}
\end{figure}

A schematic diagram of the overall instrument is shown in
Figure~\ref{fig:bulkschematic}. Light entering the instrument is split
using dichroics between the science beam combiner and the fringe
tracker. The dichroics form part of a ``beam switchyard'' which 
feeds light into the main beam combiner. This is a 4-way pupil-plane
design using beam splitters 
which produces 4 outputs, each being the superposition of all 4
input beams, i.e. interference patterns corresponding to all 6
possible baselines are present in each output. The fringes are ``fluffed out''
i.e. each output beam is either all light or all dark depending on
the phases of the fringes. Path modulators serve to 
rapidly scan the phase of the input beams such that a
temporally-modulated intensity is observed, with each of the 6 fringe
patterns appearing
at a separate frequency. The collimated outputs from the combiner are
focused onto single-mode fibres, serving to spatially filter the
fringes and also to inject light into the input ``slit'' of a cooled spectrograph. The
spectrograph illuminates a detector which 
is read out synchronously with the fringe modulation to
produce a spectro-interferogram. This can be processed to yield fringe
amplitude and phase information for all 6 baselines 
at a large number of wavelengths simultaneously.

This design is described further below in terms of a system analysis
into a number of different functions:
\begin{multicols}{2}
\begin{itemize}
\setlength{\itemsep}{0pt}
\item Atmospheric dispersion compensator modules;
\item Alignment/calibration module;
\item Beam switchyard/Fringe tracker;
\item Science beam combiner;
\item Fast path modulators
\item Beam injection module;
\item Spatial filtering;
\item Spectrograph;
\item Spectral calibration module;
\item Detector;
\item Control software;
\end{itemize}  
\end{multicols}
It can be seen that there is a large amount of commonality with the IO
solution, so in many cases the reader will be referred to the IO
sections for further detail (see section \ref{sect:conceptIO}). The
only module in the IO concept for which there is no equivalent in the
BO concept is the polarisation control module: the BO solution
requires no compensation for instrumental birefringence.  The fast
path modulators are present only in the BO concept as path modulation
is done statically in the IO combiners.

The functionality of the beam switchyard is similar to that in the IO
beam combiner, except where a 6-beam concept is implemented. With 6
input beams, the science beam combiner in the BO concept remains a
4-way design, but the switchyard serves to select subsets of the input
beams to feed to the combiner. Interferometric data are accumulated
with different subsets of input beams, with rapid (timescales of under
a minute) switching between subsets. This allows sampling of all
possible baselines and closure phases in a short period. 
As a result, the switchyard needs to be designed for rapid
reconfiguration. 

The optical arrangement which allows all 4 input beams to be
superimposed using beam splitters is shown in the box labelled ``beam
combiner'' in Figure~\ref{fig:bulkschematic}. The design uses
custom-designed coatings at low angles of incidence for high
efficiency and low polarisation sensitivity, and components are
connected using contacted-optics technology to achieve exceptional
path length and alignment stability.  The wide-band nature of the
coatings means that only one combiner is needed to cover the J, H and
K bands. The design is based on a working prototype in Cambridge which
has demonstrated the required high throughput and stability. The work
in the phase-A study would mainly involve looking at scaling issues
for the larger VLTI beams.

The optical path difference between beams is modulated at rates of
several hundred Hertz using mirrors mounted on piezoelectric
actuators. Laboratory tests have shown that the scanning of these
mirrors can achieve the required accuracy and repeatability by appropriately
controlling the harmonics of the drive waveform. Studies thus far have
used feedback from direct measurements of the resulting modulation
using a metrology laser, but injecting laser metrology to all
the modulators could be problematic within the space
envelope. Alternative options for performing the calibration of the
drive waveform include use of capacitive or strain-gauge sensors.

The injection of light into fibres in the BO design occurs after beam
combination, so the phase and birefringence properties of the fibres
are of little concern. If suitable fibres (e.g. chalcogenide photonic crystal
fibres) can be procured, then only a single set of fibres may be needed to
cover all wavebands. The beam injection technology described for the IO
combiner would be appropriate for application in the BO concept.

The detector and spectrograph layout and requirements in the BO
concept are similar to that for the IO concept. One difference is that
the BO spectrograph has fewer outputs which are read out more often
than the IO spectrograph, but the overall pixel rate is similar.
 
\subsection{Fringe tracker}

A fringe tracker performs a similar role within an interferometer as
an adaptive optics (AO) system performs within a single telescope. It
measures wavefront errors due to the atmosphere and instrument (in the
case of a fringe tracker, these OPD errors are the differences between
telescopes of the amplitudes of their respective Zernike ``piston''
modes) and corrects these errors in real time. Like an AO system, a
fringe tracker requires a bright reference object to sense the
wavefront errors, and also like an AO system there are strong
fundamental limits to how faint this reference object is allowed to be
before the fringe tracker fails to operate.  If the fringe tracker
does not work, little or no science can be done, and so the range of
science targets accessible to the instrument is typically most
strongly limited by the performance of the fringe tracker, especially
its magnitude limit.  Because of this intimate relationship between
fringe tracker performance and the science performance of the system
as a whole, VSI will incorporate its own fringe tracking
subsystem. The combined system of fringe tracker and science beam
combiner will be optimised as a whole to meet the science
requirements.

In order to fulfill the science goals of high-throughput acquisition
of data, high-SNR high-spectral-resolution imaging, and imaging of
complex faint objects, the fringe tracker will operate in a minimum of
three distinct modes defined as follows:

\paragraph{Fringe acquisition} 

This is a mode in which a finite region of OPD space is scanned in
order to find the fringe coherence envelope in the presence of
atmospheric and instrumental delay uncertainties. This mode will
typically use either a continuous or stepped scan of the delay lines
together with a group-delay algorithm to efficiently detect the presence of
fringes over a region of delay space set by the coherence length of a
single spectral channel.

\paragraph{Hardware phase tracking} 

This is a mode in which the delay errors are actively compensated at
high speed so that the level of stability of the science combiner
fringes is sufficient to allow on-chip integration of the fringe
signal over periods of many atmospheric coherence times. Typically
hardware phase tracking requires a high SNR fringe phase measurement
to be made in a short integration time to allow the high-precision
(typically of order $\lambda/20$) correction required. This means that
a bright ($m_H<10$) reference object is required, but providing such a
reference is available, then the long coherent integration times
afforded on the science combiner allow high-spectral-resolution
measurements to be made relatively rapidly.

\paragraph{Hardware coherencing} 

This is a mode in which the delay errors are compensated in real time
but at a lower speed, with sufficient precision to ensure that the
loss in fringe contrast due to temporal coherence effects is small
over incoherent integration periods of many minutes. Using group-delay
tracking methods on spectrally-dispersed fringes
\cite{2005MNRAS.357..656B}, the fringes can be tracked using reference
stars at least 2.5 magnitudes fainter than are usable with
phase-tracking methods \cite{1989Buscher}.  In addition, in the
presence of short-term Strehl ``dropouts'' and phase branch point
effects due to imperfect AO correction, group delay methods are
considerably more robust. Thus the hardware coherencing mode will be
most beneficial for science on faint targets and/or in moderate to
poor seeing. In this mode, the science combiner will need to be read
out at rates comparable to the atmospheric coherence time, rather than
the longer integration times afforded by the phase-tracking mode.

\begin{figure}[t]
  \centering
  \includegraphics[width=0.7\hsize]{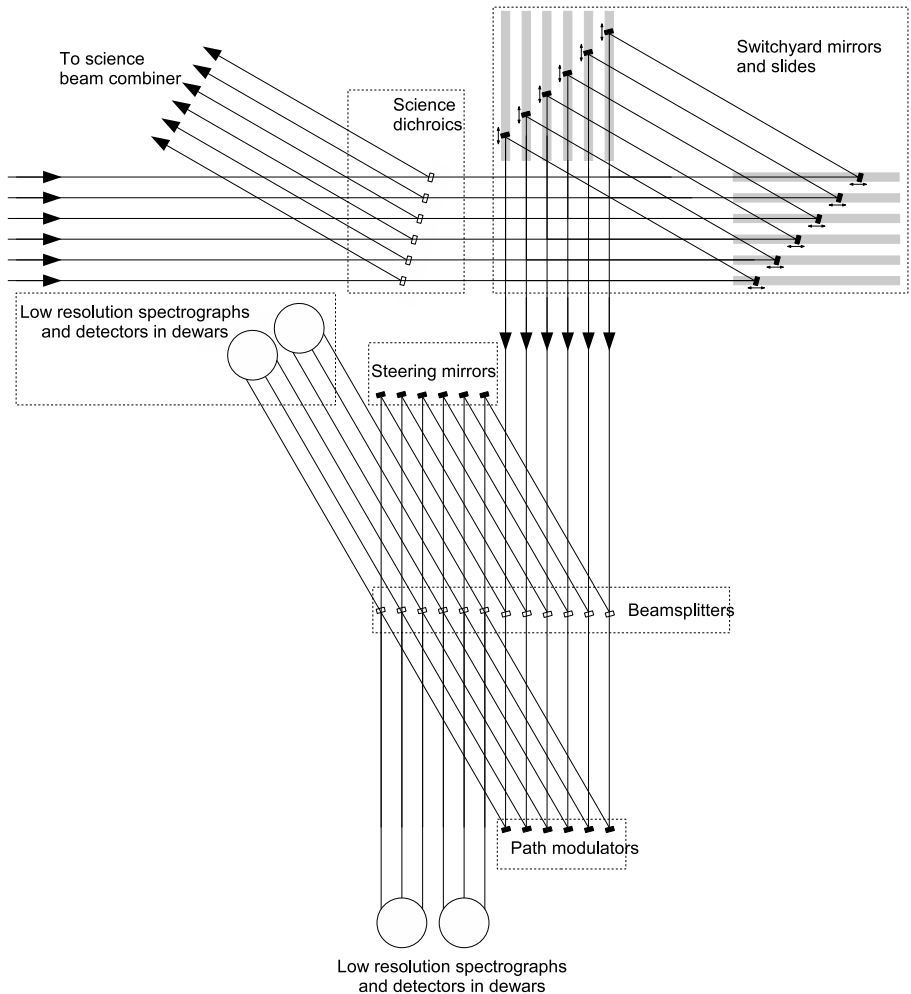}
  \caption{Conceptual layout of the fringe tracker optics}
\label{fig:tracker}
\end{figure}

A conceptual layout for the fringe tracker is shown in
Fig.~\ref{fig:tracker}. The subcomponents of the fringe tracker which are
described have been separated into:
\begin{multicols}{2}
\begin{itemize}
\setlength{\itemsep}{0pt}
\item Dichroics
\item Beam switchyard;
\item Fringe-tracking beam combiner;
\item Fast path modulators
\item Low-resolution spectrograph and detector;
\item Control system;
\item OPD corrector;
\end{itemize}
\end{multicols}
A complete optical and mechanical design of the fringe-tracking
combiner optics is needed to ensure that the combiner can be fitted
within the space envelope at the VLTI.

A key parameter of the science case for the instrument as a whole will
be the list of which potential science targets can be observed in a
high-spectral-resolution phase-tracking mode and which can be observed
in a faint-object coherencing mode.  Detailed calculations of the
magnitude limits for both these modes will feed into the determination
of these potential target lists.

The maximum coherent integration time, and hence SNR, for the science
combiner when the fringe combiner is in phase-tracking mode will be
strongly limited by drifts in the relative OPD as measured by the
fringe tracker and as seen by the science beam combiner. Various
methods of ``tying together'' the beam combiners using either
mechanical means or laser metrology will be evaluated to determine the
practical limits to achieving long coherent integrations and the most
cost-effective means for achieving these.

The real-time component of the fringe tracker derives adjustments to
the OPD from the pixel data stream arriving from the detectors.  In
both the coherencing or phase-tracking modes the fringe amplitude and
phase are derived for each of the 5 spectral sub-bands at rates of up
to several hundred Hz. Typically 4 fringe samples need to be taken
synchronously with the OPD modulation in order to derive the fringe
parameters, so that 20 pixels in total need to be processed per
coherent integration per baseline being tracked. For 5 baselines being
tracked at a sample rate of 500Hz, the data rate is 50ksamples/sec.
This is not a demanding computational task, and so specialised
computing hardware (e.g. DSP arrays) is not necessary --- it should be
possible to perform all the real-time computation on a single
Pentium-class processor running a hard real-time operating system.

The fringe-tracking computer filters the derived OPD errors and sends
the appropriate signals to a two-stage OPD correction system
consisting of a mirror mounted on a fast piezo-electric stage
connected which takes out rapid pathlength fluctuations, and, error
signals sent to the delay line signals to correct long-term
large-amplitude OPD errors.

\section{Conclusion}

VSI is proposed as second generation VLTI instrument providing the ESO
community with the capability of performing image synthesis at
milli-arcsecond angular resolution.  VSI provides the VLTI with an
instrument capable of combining 4 telescopes in a baseline version and
optionally up to 6 telescopes in the near-infrared spectral domain
with moderate to high spectral resolution. The instrument contains its
own fringe tracker in order to relax the constraints onto the VLTI
infrastructure.  Two solutions for the science beam combiner were
identified one based on integrated optics and another on bulk optics:
\begin{itemize}
\item The integrated optics science beam combiner solution has been
  validated with astrophysical results at high performance on IOTA and
  VLTI.  It emphasizes the maintainability of the instrument (apart
  from the injection devices, there are no degrees of freedom left for
  the beam combination); it is well suited to feed a conventional
  infrared spectrograph; enhancing the number of telescopes from 4 to
  6 just requires a different IO device which can be fed by different
  fibers and the duplication of 2 more injection modules.
\item The bulk optics science beam combiner solution has an emphasis
  on the commonality with the integrated optics solution. It is based
  on a 4-way working prototype in Cambridge. Although having a larger
  number of degrees of freedom, it has high optical throughput and
  interferometric contrast. To work with 6 telescopes, the beam
  combiner requires fast switching optics in order to select subsets
  of the input beams to feed the 4-way beam combiner. The 4 outputs of
  the beam combiner are injected in fibers to feed a similar
  spectrograph as the one designed for the IO solution.
\end{itemize}
These two solutions are inherited from the two concepts merged. One of
the goals of the Phase A study is to define which of them will be used
by VSI.  The phase A starts in June 2006 and will last one year until
the study will be reviewed by ESO who will take a final decision.


\bibliography{vsi}   
\bibliographystyle{spiebib}   

\end{document}